\documentclass[twocolumn,showpacs,pra,aps]{revtex4-1}
\usepackage{graphicx}
\usepackage{epstopdf}
\usepackage{amsmath}
\usepackage{amssymb}
\usepackage{epsfig}
\usepackage{amsthm}
\usepackage{color}
\begin{document}
\title{Bell Inequalities for Quantum Optical Fields}
\author{Marek \. Zukowski}
\author{Marcin Wie\'sniak}
\author{Wies\l aw Laskowski}
\affiliation{Institute of Theoretical Physics and Astrophysics, University of Gda\'{n}sk, 80-952 Gda\'{n}sk, Poland}

\begin{abstract}
We show that the ``practical" Bell inequalities, which use intensities as the observed variables,  commonly used in quantum optics and widely accepted in the community, suffer from an inherent loophole, which severely limits the range of local hidden variable theories of light, which are invalidated by their violation. We present alternative  inequalities which do not suffer from any (theoretical) loophole. The new inequalities use redefined correlation functions, which involve averaged  products of  local rates  rather than intensities. Surprisingly, the new inequalities detect entanglement in situations in which the ``practical" ones fail. Thus, we have two for the price on one:  full consistency with Bell's Theorem, and better device-independent detection of entanglement.
\end{abstract}
\maketitle

%------------------------------------------------------------
\section{Introduction}
%------------------------------------------------------------
Since the first derivation of Bell inequality based, except for ``freedom",  only on the assumption of locality and (a form of) realism (or hidden causality, states, variables) which one can find in the CHSH paper \cite{CHSH}, many other ones were derived. They either apply to different number of settings, or different number of systems and observers, or different observables (for a recent review of this ``zoo" see Ref. \cite{WEHNER}). Here we would like to study the case of Bell inequalities which are specially tailored for general quantum optical situations, especially the cases for fields of varying intensities (essentially, undefined photon numbers).  The case of strongly driven type-II parametric down conversion (with multiple pair emissions), which we shall call bright squeezed vacuum will be our working example here. But, obviously, multitude of other examples can be given. We use in the title the phrase ``quantum optical fields" to stress that the inequalities we study do not assume a fixed (overall) photon number.

Such inequalities play an important role in singling out non-classical phenomena in the optical domain. E.g. with the use of such inequalities
Bell's theorem for approximate EPR states was finally given in  \cite{BANASZEK}. The idea was to use different observables than the ones discussed by EPR . In \cite{BANASZEK} observables with singular Wigner representations were used (essentially, photon number parity operators measured after a displacement). 
Still, in general quantum optics there is an obvious demand for device-independent entanglement indicators (witnesses) and therefore Bell inequalities for variables straightforwardly related to the usual measured observables.  We shall propose such. But first we shall present the principal work which addressed the problem, and contains a kind of mainstream approach.

\subsection{Standard approach to Bell Inequalities for quantum Optics}
We shall review here the derivation by Reid and Walls \cite{REID}, which set the standard in the field, for many years.
Two observers measure intensities of light at outputs of some local interferometric devices. It is customary to think of them as outputs of polarization analyzers.
We shall follow this picture here, as our aim is to convey to the reader the general ideas. Let us concentrate on the derivation of Clauser-Horne-like \cite{CH} inequalities, as they are often thought of as the most basic ones.

The approach of Ref. \cite{REID} would involve the following elements of the hidden variable description. We have intensities potentially measurable by Alice in the two outputs of he analyzer $I_{A_\pm}(\theta, \lambda)$, which are non-negative (an extension of the approach of  to this regime  \cite{REID} was given in \cite{ZUK-1989}) and in principle unbounded. For Bob we have $I_{B_\pm}(\phi, \lambda)$. Here, $\theta$ and $\phi$ symbolize the local settings of the polarization analyzers (the can be replaced, if needed,  by description characteristic for elliptic polarization measurements, like local Bloch vectors, etc.), controllable by the observers, and $\lambda$ stands for  hidden variables (hidden causes not present in the quantum description). The hidden variables are assumed to be distributed with a probability density $\rho(\lambda)$.  Up to this point all that seems to be perfectly in line with the usual thinking behind Bell inequalities

Had the above been the only set of assumptions about local hidden variable description, one would not expect any loopholes in the resulting inequalities. However, next in Ref. \cite{REID} comes a remark that the ``total intensity through each polarized is  written" 
\begin{eqnarray}\label{ASSUMP1}
&I_A(\lambda)=I_{A_+}(\theta, \lambda)+I_{A_-}(\theta, \lambda), &\nonumber \\
&I_B(\lambda)=I_{B_+}(\phi, \lambda)+I_{B_-}(\phi, \lambda).&
\end{eqnarray}
It is argued that $I_A(\lambda)$ and $I_B(\lambda)$ are in principle independent of $\theta$ and $\phi$, and that they correspond to local intensities with polarizers removed. All this is an {\em additional} assumption on the form of hidden variable theories, which does not have to hold generally in all possible hidden variable descriptions.  E.g. polarizer in such descriptions could "enhance" for some values of $\lambda$ the total intensity registered behind it, and for some lower it. Let us present  an old argument. Note that in classical optics if one has two polarization filters on the way of a beam, and the first one transmits  the horizontal light only and the second one only the vertical one, then no light exits such two polarizer-system. However insertion, between the two, of a polarization filter selecting diagonally polarized light  results in an output intensity which is $1/4$ of the original one. Thus, the presence of  polarizers does not always reduce intensity. One could here introduce a version of "no enhancement" assumption, namely $I_A(\lambda)\geq I_{A_+}(\theta, \lambda)+I_{A_-}(\theta, \lambda), $  etc., but this also is an arbitrary constraint on hidden variable models.

The additional argumentation against such an assumption can come directly from quantum optics. Take as an example (bright) squeezed vacuum. In such a state the intensity is undefined.  {However, one should not here that the condition perfectly holds for classical stochastic theory of light, and this could have been the intuition that led the authors of \cite{REID} to adopt it. Still, the condition does {\em not} hold in the case ``stochastic electrodynamics \cite{MARSHALL}}.  Even if we argue that the condition reflects energy conservation, this can be challenged by the fact that the field {\em plus the system of detectors} is in fact an open system. Thus the assumption seems not so much justified, afterall. One of the aims of this work is to formulate quantum optical Bell theorem for systems of varying intensities which is free of such assumptions.

The additional assumption (\ref{ASSUMP1}) allows one to construct local hidden variable correlation functions which are constrained by CHSH-like \cite{CHSH}
inequalities.

\begin{equation}
\label{REIDEXP}
E(\theta, \phi)=\frac{\langle(I_{A_+}(\theta)-I_{A_-}(\theta))(I_{B_+}(\phi)-I_{B_-}(\phi))\rangle}{\langle I_{A}I_{B}\rangle}
\end{equation}
where 
\begin{eqnarray}
&\langle(I_{A_+}(\theta)-I_{A_-}(\theta))(I_{B_+}(\phi)-I_{B_-}(\phi))\rangle& \nonumber\\
&= \int \rho(\lambda)(I_{A_+}(\theta, \lambda)-I_{A_-}(\theta, \lambda))(I_{B_+}(\phi, \lambda)-I_{B_-}(\phi, \lambda)), &\nonumber \\
\end{eqnarray}
while
\begin{eqnarray}
&\langle  I_{A}I_{B}\rangle= \int d\lambda\rho(\lambda)I_{A}(\lambda))I_{B}( \lambda). &
\end{eqnarray}
with all that one quickly arrives at the usual form of the CHSH-like inequality:
\begin{equation}
\label{CHSH}
|E(\theta,\phi)+ E(\theta,\phi')+E(\theta',\phi)-E(\theta',\phi')|\leq 2,
\end{equation}
for details see Ref. \cite{REID}. This possible {\em only} due to the fact that $D=\langle  I_{A}I_{B}\rangle$ is the common denominator in all correlation functions entering the CHSH-like inequality.  Thus assumption (\ref{ASSUMP1}) is crucial in the derivation.

In the case of Clauser-Horne-like inequalities one proceeds similarly. However, in the case of considerations in Ref. \cite{REID} the no-enhancement assumption is used instead of (\ref{ASSUMP1}). Namely  that there exists  variables, which are {\em independent} of the local setting (and the remote one...) $I_A(\lambda)$  and $I_B(\lambda)$ such that 
 \begin{eqnarray}\label{ASSUMP2}
&I_A(\lambda)\geq I_{A_+}(\theta, \lambda), &\nonumber \\
&I_B(\lambda)\geq I_{B_+}(\phi, \lambda).&
\end{eqnarray}
This allows to introduce functions
\begin{equation}
G(\theta, \phi)=\int d\lambda\rho(\lambda) I_{A_+}(\theta, \lambda)I_{B_+}(\phi, \lambda),
\end{equation}
and 
\begin{eqnarray}
r_A(\theta)=\int d\lambda \rho(\lambda)I_B(\lambda)I_{A_+}(\theta, \lambda), \nonumber \\
r_B(\phi)=\int d\lambda \rho(\lambda)I_A(\lambda) I_{B_+}(\theta, \lambda). \nonumber \\
\end{eqnarray}
The variables $I_A(\lambda)$  and $I_B(\lambda)$ are treated  as intensities (potentially) measured by the local observers ``without the polarizers present". The additional assumption (\ref{ASSUMP2})  is not in conflict with (\ref{ASSUMP1}), as it logically  follows from it. Still, it wides the class of hidden variable theories under consideration. Nevertheless, is prone to the same criticism as   (\ref{ASSUMP1}).

To derive the actual inequality one can use the Clauser-Horne lemma: for real numbers  $0\leq x, x' \leq X$ and $0\leq y, y' \leq Y$ one has
\begin{equation}
xy+xy'+x'y-x'y'-xY-Xy\leq 0,
\end{equation}
(in Ref \cite{REID} there is a typo in the equivalent inequality).
Thus 
\begin{eqnarray}
 I_{A_+}(\theta, \lambda)I_{B_+}(\phi, \lambda)+ I_{A_+}(\theta, \lambda)I_{B_+}(\phi', \lambda)\nonumber \\
+ I_{A_+}(\theta', \lambda)I_{B_+}(\phi, \lambda) - I_{A_+}(\theta', \lambda)I_{B_+}(\phi', \lambda)\nonumber \\
- I_{A_+}(\theta, \lambda)I_{B}(\lambda) - I_{A}( \lambda)I_{B_+}(\phi, \lambda)\leq0.
\end{eqnarray}
This after averaging allows one to formulate a Clauser-Horne-like inequality
\begin{eqnarray}
 &G(\theta, \phi,)+ G(\theta,\phi')
+ G(\theta',\phi)& \nonumber \\
&- G(\theta', \phi')
- r_{A}(\theta) - r_{B}(\phi)\leq0.&
\end{eqnarray}

The above  inequalities. like most of other Bell inequalities, e.g. the original CHSH ones,  can be used in the analysis of the formalism of quantum optics and its predictions aimed at proving  the impossibility of having a local realistic model of all that.
Both types of inequalities can be violated in quantum optical experiments involving quantum entanglement.  As a matter of fact, in very many quantum optical experiments  the inequalities were used in analysis of  results. Still, the validity of such results (theoretical and experimental) is limited by the fact that we use the additional assumptions, already at the level of {\em derivation} of the inequalities. Other loopholes (detection loophole, locality loophole), which are due to experimental imperfections, can pop up in the analysis of the experimental results, however these are not a trait of the above inequalities. 

\section{New Approach}

We shall show that one can derive Bell inequalities for quantum optics of a different kind, but still being generalizations of the CHSH ones and CH ones, which are free of the above additional assumptions and thus loophole free (the loopholes can now appear only in the methods of execution of the actual experiments). We shall additionally argue that they are usually more strongly violated than the ones presented above.

The inequalities will use ``rates" instead of intensities.
We define them as 
\begin{eqnarray}&R_{A_{\pm}}(\theta)=\frac{I_{A_\pm}(\theta)}{I_{A}(\theta)}, &\\
&R_{B_{\pm}}(\phi)=\frac{I_{B_\pm}(\phi)}{I_{B}(\phi)}.& \\ \nonumber 
\end{eqnarray}
 and from the operational point of view  they show the relative distribution of the measured intensities at the two outputs (we still use the two output working examples, but all that can be generalized to more complicated cases) of the local polarizer. We do not assume that  the total intensity is independent of the local setting. Instead we define the following variables 
\begin{eqnarray}
&{I_{A}(\theta)}=\sum_{i=\pm }{I_{A_i}(\theta)}, &\nonumber \\
&I_{B}(\phi)=\sum_{i=\pm }I_{B_i}(\phi).& \\
\end{eqnarray}
Of course in quantum theory averages  both quantities do not depend on on the local settings of the polarizes (well, in the experiment one could expect some variation due to imperfections affecting losses), but we introduce such a description having in mind the hidden variable theories for which ``anything goes".

Let us begin with a derivation of Clauser-Horne-like inequalities.

\subsection{CH-like inequalities}
In this derivation we shall use a hidden variable model of the rates. Our only basic tool will be $I_{A_\pm}(\theta, \lambda)$ and $I_{B_\pm}(\phi, \lambda)$.
In the hidden variable description we shall assume existence of values for all possible settings of the following variables:
\begin{equation}
R_{A_+}(\theta, \lambda)=\frac{I_{A_+}(\theta, \lambda)}{I_{B}(\theta, \lambda)}
\end{equation}
where $I_{A}(\theta, \lambda)=\sum_{i=\pm} I_{A_i}(\theta,\lambda), $
and
\begin{equation}
R_{B_+}(\phi, \lambda)=\frac{I_{B_+}(\phi, \lambda)}{I_{A}(\phi, \lambda)},
\end{equation}
with $I_{B}(\phi, \lambda)=\sum_{i=\pm} I_{B_i}(\phi,\lambda). $
Whenever the given  denominator {\em vanishes}, for certain $\lambda$ and certain local setting, we additionally define that for the given (hidden) rate one has $R_{J_\pm} (\alpha,\lambda)=0$, where $J=A$ or $B$, and $\alpha=\theta, \phi$, respectively.

Obviously one has $0 \leq R_{J_+}(\alpha, \lambda)\leq1$. Thus the Clauser-Horne lemma can be used to get 
\begin{eqnarray}
&-1\leq R_{A_+}(\theta, \lambda)R_{B_+}(\phi, \lambda)+ R_{A_+}(\theta, \lambda)R_{B_+}(\phi', \lambda)&\nonumber \\
&+ R_{A_+}(\theta', \lambda)R_{B_+}(\phi, \lambda) - R_{A_+}(\theta', \lambda)R_{B_+}(\phi', \lambda)&\nonumber \\
&- R_{A_+}(\theta, \lambda) - R_{B_+}(\phi, \lambda)\leq0.&
\end{eqnarray}
This after averaging over $\rho(\lambda)$ leads to
\begin{eqnarray}
&-1\leq K(\theta, \phi)+ K(\theta,\phi')+ K(\theta', \phi) - K(\theta', \phi')&\nonumber \\
&- S_{A_+}(\theta, \lambda) - S_{B_+}(\phi, \lambda)\leq0,&
\end{eqnarray}
where 
\begin{equation}
K(\theta, \phi)=\int d\lambda\rho(\lambda) R_{A_+}(\theta, \lambda)R_{B_+}(\phi, \lambda),
\end{equation}
and 
 \begin{equation}
S_{J_+}(\alpha)=\int d\lambda\rho(\lambda) R_{J_+}(\alpha, \lambda).
\end{equation} 
This is the new form of a loophole free Clauser-Horne type inequality for quantum optical experiments. The loopholes are moved to the experimental imperfections, like inefficient detection, etc.

The reader might be worried that we divide here by the intensity, and how should look like the quantum definition for equivalent expression. Of course this depends on the model of detection used, which defines the intensity operators at the exists of the local devices (polarizers). to illustrate this we shall use 
the simplest model: the photon number operator. 

As we have four outputs, and an example involving polarizations, let us introduce annihilation (and creation) operators $a_H$, $a_V$, $b_H$ and $b_V$ (creation operators would have a dagger, $\dagger$). Here $H$ and $V$ stand for horizontal and vertical polarizations, and $a, b$ denote propagation beams, towards Alice and Bob, respectively. The operators for arbitrarily oriented (linear) polarization analyzers are:
  \begin{eqnarray}
\cos \theta a_H+\sin\theta a_V= a_{\theta},\\
-\sin \theta a_H+\cos\theta a_V= a_{\theta^\perp},\\
\cos \phi b_H+\sin\phi b_V= b_{\phi},\\
-\sin \phi b_H+\cos\phi b_V= b_{\phi^\perp}.\\ \nonumber
\end{eqnarray}
Let us assume the the operators with $\perp$ denote  $-$ exit of the local polarizers, and the other ones  $+$ output. Under the chosen intensity model
given by 
\begin{eqnarray}
\hat{I}_{J_+} = c_\alpha^\dagger c_\alpha,\\
\hat{I}_{J_-} = c_\alpha^\dagger c_{\alpha^\perp},
\end{eqnarray}
where $c=a,b$, $\alpha=\theta, \phi$ for $J=A,B$, respectively. the quantum versions of the functions in the inequality read:
\begin{equation}\label{QUANTUM}
K(\theta, \phi)_Q=Tr [\varrho  \hat{R}_{A_+}(\theta)\hat{R}_{B_+}(\phi)],
\end{equation}
and 
 \begin{equation}
S_{J_+}(\alpha)_Q=Tr[ \varrho  \hat{R}_{J_+}(\alpha)],
\end{equation} 
where $\varrho$ is the quantum state, and 
$ \hat{R}_{J_+}(\alpha)$ are given by:
\begin{equation} 
 \hat{R}_{J_+}(\alpha)= \hat{\Pi}_J^{\perp 0}\frac{\hat{I}_{J_+}}{\sum_{i=\pm}\hat{I}_{J_\pm}} \hat{\Pi}_J^{\perp0}.
\end{equation}
The  operators $\hat{\Pi}_J^{\perp 0}$ are projectors  into the subspace, of the Fock space, of the modes described by annihilation operators $c_H$ and $c_V$, which does not contain vacuum. 

In case the notation seems a bit too dense, let us write an example of such an operator explicitly.  E.g., the operator  $\hat{R}_{a_+}(\theta)$ acts in the {\em joint} Fock space of modes $a_{\theta}$ {\em and} $a_{\theta^\perp}$, and reads
\begin{eqnarray} 
 \hat{R}_{A_+}(\theta)
= \hat{\Pi}_A^{\perp 0}\frac{a^\dagger_\theta a_\theta}{a^\dagger_\theta a_\theta+a^\dagger_{\theta_\perp} a_{\theta_\perp}}  \hat{\Pi}_A^{\perp 0}
\end{eqnarray}
with
\begin{eqnarray}
\hat{\Pi}_A^{\perp 0}= \hat{I}_a - |0,0\rangle_{aa}\langle0,0|,
\end{eqnarray}
where   $\hat{I}_a$ is the identity operator in the space, and $|0,0\rangle_{a}$ is the vacuum of the space satisfying $a_\theta|0,0\rangle_{a}=a_{\theta^\perp}|0,0\rangle_{a}=0.$

As in quantum optics violations of Bell inequalities of the type presented in \cite{REID} usually  decreases with the growing number of photons, see e.g. \cite{LASK}, the inequalities
 introduced here might be less prone to this effect, because their form warrants that  that terms with higher photon numbers contribute less to correlation functions $K(\theta, \phi)_Q$ and to the local averages $S_{J_+}(\alpha)_Q$.
Thus we shall benefit from their use twice. They are not only formulated without additional assumptions (which is a kind of built in loophole), but also we shall gain on the sensitivity of the inequalities in detecting non-classicality. This will be shown in the examples of the next Section.
But before this we shall derive CHSH-like Bell inequalities for quantum optics based on similar concepts.

\subsection{CHSH-like inequalities}
The CHSH-like inequalities can be derived  using the same basic hidden variable objects  $I_{A_\pm}(\theta, \lambda)$ and $I_{B_\pm}(\phi, \lambda)$, and their functions  
$R_{A_+}(\theta, \lambda)$ and
$R_{B_+}(\phi, \lambda)$, plus additional ones 
\begin{equation}
R_{A_-}(\theta, \lambda)=\frac{I_{A_-}(\theta, \lambda)}{I_{A}(\theta, \lambda)}
\end{equation}
where $I_{A}(\theta, \lambda)=\sum_{i=\pm} I_{A_i}(\theta,\lambda), $
and
\begin{equation}
R_{B_-}(\phi, \lambda)=\frac{I_{A_-}(\phi, \lambda)}{I_{A}(\phi, \lambda)}.
\end{equation}
Recall the all $R$ variables equal zero for zero total local intensity.
Obviously this can be done 
by introducing correlation functions
\begin{eqnarray}
F(\theta,\phi)=\int d\lambda \rho(\lambda)[R_{A_+}(\theta, \lambda)-R_{A_-}(\theta, \lambda)] \nonumber \\
\times [R_{B_+}(\phi, \lambda)-R_{B_-}(\phi, \lambda)].
\end{eqnarray}
As $ -1\leq R_{J_+}(\alpha, \lambda)-R_{J_-}(\alpha, \lambda)\leq1$, one can proceed in the usual way to derive:
\begin{equation}
|F(\theta,\phi)+ F(\theta,\phi')+F(\theta',\phi)-F(\theta',\phi')|\leq 2.
\end{equation}
while this is a correct inequality, its usefulness for quantum optics is highly limited. This is because in most Bell experiments the produced 
states are such that the vacuum components dominate. Especially this is the case for parametric down conversion, see e.g. \cite{PAN}.
As a result the values of the correlation functions are very low, and the left hand side cannot breach the bound of 2.
Still such inequalities could be of use in the case of event ready experiments, see \cite{ZZHE}.

To devise CHSH inequalities which are immune to this deficiency, one can proceed as follows.
One can introduce new rate functions which behave differently when the intensity is {\em zero}.
Simply one can redefine the (arbitrary) value of $R_{A_+}(\theta, \lambda)$ and
$R_{B_+}(\phi, \lambda)$ for zero total intensity, and put is as $1$. No change is required for $R_{J_-}(\alpha, \lambda)$.
We can denote the redefined rate functions by $R_{A_+}'(\theta, \lambda)$ and
$R_{B_+}'(\phi, \lambda)$. 
Again we have  $$ -1\leq R_{J_+}'(\alpha, \lambda)-R_{J_-}(\alpha, \lambda)\leq1.$$
With the new variables the hidden variable model for the correlation function reads:
\begin{eqnarray}
\label{NEWCORR}
C(\theta,\phi)=\int d\lambda \rho(\lambda)[R_{A_+}'(\theta, \lambda)-R_{A_-}(\theta, \lambda)] \nonumber \\
\times [R_{B_+}'(\phi, \lambda)-R_{B_-}(\phi, \lambda)].
\end{eqnarray}
This definition does not suffer from the mentioned deficiency, because in the case of vacuum the values $1$ pop up.
We have a modified inequality:
\begin{equation}
\label{CHSHC}
|C(\theta,\phi)+ C(\theta,\phi')+C(\theta',\phi)-C(\theta',\phi')|\leq 2.
\end{equation}
In the case of vacuum state the modified  inequality, following from the above definitions, 
is {\em saturated}, while for some other quantum state it can be violated, see further down.

We still have to give a quantum formula for $C(\theta,\phi)_Q$. To this end one must define  the operator $ R_{J_+}'(\alpha)$. This will be again done for the trivial model of intensity in terms of the number operator.
\begin{equation}
\label{RPA} 
\hat{R}_{A_+}'(\theta)=\hat{R}_{A_+}(\theta)+|0,0\rangle_{aa}\langle0,0|
\end{equation}
and
\begin{eqnarray}
\label{RPB} 
\hat{R}_{B_+}'(\phi)
=\hat{R}_{B_+}(\phi)+|0,0\rangle_{bb}\langle0,0|,
\end{eqnarray}
where $|0,0\rangle_{b}$
is the vacuum for the $b$ modes.

\section{Examples}

Surprisingly the new inequalities, not only are free of an inbuilt loophole (that is the additional condition), but one can easily find examples of quantum optical states with violate them more robustly than the old-type inequalities. We shall consider here our working example  of the four mode (bright) squeezed vacuum.  The state reads
\begin{eqnarray}
\label{BSV}
|BSV\rangle=\frac{1}{\cosh^2{\Gamma}}\sum_{n=0}^{\infty} \sqrt{n+1}
\tanh^{n}{\Gamma} |\psi^{(n)}_- \rangle,
\end{eqnarray}
where
\begin{eqnarray}
&|\psi^{(n)}_-\rangle&\nonumber \\&=\frac{1}{\sqrt{n+1}} \sum_{m=0}^n
(-1)^m|n-m\rangle_{a_H} |m\rangle_{a_V} |m \rangle_{b_H} |n-m\rangle_{b_V},&\nonumber \\
\end{eqnarray}
where $a$ and $b$ refer to the two directions along which the photon pairs
are emitted,  $H/V$ denote horizontal/vertical polarization, and $\Gamma$
represents an amplification gain.
The state is rotationally invariant, i.e., all probabilities are dependent only on $\theta-\psi$, and this can also holds for  elliptic polarizations.. 

First, let us compute the value of the CHSH expression in fashion of Ref. \cite{REID}, according to formula (\ref{REIDEXP}). For every Fock component of state (\ref{BSV}) with $2n$ photons , the correlation function is a cosine function with one of its minima at $\theta-\phi=0$ and amplitudes $\frac{1}{3}(2n+n^2)$. Thus, the numerator in Eq. (\ref{REIDEXP}), summed over $n>0$ with weights $\frac{(n+1)\tanh^{2n}\Gamma}{\cosh^4\Gamma}$, reads
\begin{eqnarray}
&&\langle(I_{A_+}(\theta)-I_{A_-}(\theta))(I_{B_+}(\phi)-I_{B_-}(\phi))\rangle\nonumber\\
&=&-2\cosh^2\Gamma\sinh^2\Gamma\cos(\theta-\phi),
\end{eqnarray}
while the denominator is equal to
\begin{eqnarray}
\langle I_{A}I_{B}\rangle
=\frac{5}{4}-2\cosh 2\Gamma+\frac{3}{4}\cosh 4\Gamma+\cosh^4\Gamma.&& \\ \nonumber
\end{eqnarray}
If we optimal settings $\theta=\pi,\theta'=\frac{3}2\pi,\phi=\frac{1}{4}\pi,$ and $\phi'=-\frac{1}{4}\pi$ , we get violations of the Bell inequality  for $\Gamma<0.4911$.

In the new loophole-free approach utilizing the ratios, for an individual $2n$-photon component we have
\begin{eqnarray}
&\langle\psi_-^{(n)}|[R_{A_+}'(\theta)-R_{A_-}(\theta)][R_{B_+}'(\phi, \lambda)-R_{B_-}(\phi, \lambda)]|\psi_-{(n)}\rangle& \nonumber\\
&=-\frac{n+2}{3n}\cos(\theta-\phi), &
\end{eqnarray}
while $\langle 0,0,0,0|[R_{A_+}'(\theta)-R_{A_-}(\theta)][R_{B_+}'(\phi, \lambda)-R_{B_-}(\phi, \lambda)]|0,0,0,0 \rangle=-1$. After averaging over Fock components with weights $(n+1)\tanh^{2n}(\Gamma)\cosh^{-4}\Gamma$, we get
\begin{eqnarray}
&C(\theta, \phi)_Q
=\frac{1}{\cosh^4\Gamma}\left[1-\frac{1}{3}\cos(\theta-\phi)\left(4\log\cosh\Gamma\right.\right.&\nonumber\\
&\left.\left.+(3+\cosh^2\Gamma)\sinh^2\Gamma\right)\right]. &
\end{eqnarray}
This put into the CHSH expression provides violation for $\Gamma<0.8866$. Not only does inequality (\ref{CHSHC}) provide a fully valid  Bell inequality for quantum  optical fields, it also extends a region of gain $\Gamma$, for which they can be violated.

\subsection{CH-type inequality}

\begin{figure}
\includegraphics[width=0.45\textwidth]{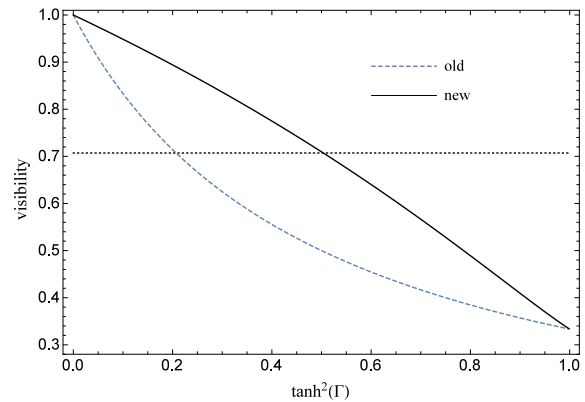}
\caption{\label{visibility} Interference contrast (visibility) for quantum prediction for $K(\theta, \phi)_Q$ (new) and  $G(\theta, \phi)_Q$ (old). The usual threshold for violation of CHSH ineqialties is given by the straight line, just to indicate the scale of improvement.  }
\end{figure}

After some algebraic manipulations the coincidence rate function $K(\theta, \phi)_Q$ (\ref{QUANTUM}) for the  bright squeezed vacuum state (\ref{BSV}) can  be found to be
\begin{eqnarray}
&K(\theta, \phi)_Q =  \frac{1}{4} \left(1 - \frac{1}{\cosh^4 \Gamma}\right)  + 
 \frac{\cos(\theta- \phi)}{96 \cosh^4 \Gamma}& \\
    &\times  \left(-13 + 12 \cosh 2 \Gamma + \cosh 4 \Gamma + 8 \ln {\cosh \Gamma}\right).& \nonumber
\end{eqnarray}
The visibility (contrast) of the interference is therefore
\begin{eqnarray}
v_{new}(\Gamma) &=& \frac{K_{max}(\theta, \phi)_Q-K_{min}(\theta, \phi)_Q}{K_{max}(\theta, \phi)_Q+K_{min}(\theta, \phi)_Q} \\&=& \frac{3 + \cosh^2{\Gamma} - 4 \frac{\ln{\cosh \Gamma}}{\sinh^2 \Gamma} }{3 + 3 \cosh^2{\Gamma}},
\end{eqnarray}
where the minimum and maximum is taken over $\phi$ and $\theta$.
In the case of the appraoch of \cite{REID} the visibility is given by  (see e.g. \cite{LASK})
\begin{equation}
v_{old}(\Gamma)=\frac{1}{1+2 \tanh^2 \Gamma}.
\end{equation}
The visibilities for rates are much more pronounced, see Fig. {\ref{visibility}}.

%CH INEQUALITIES, PLEASE.

\section{Extension of the method}

We have also compared the thresholds for these two approaches when adopted to the chained inequalities with $L$ settings per side. Such inequalities (for the new approach) read
\begin{eqnarray}
\label{CHSHC}
&|C(\theta_1,\phi_1)+ C(\theta_2,\phi_1)+C(\theta_2,\phi_2)&\nonumber \\ &+ ... + C(\theta_L,\phi_L)-C(\theta_1,\phi_L)|\leq 2(L-1).&
\end{eqnarray}
This was done for equal spacings of the settings $\theta_i=\frac{(2i-1)\pi}{4L}$ and $\phi_i=\frac{2\pi i}{4L}$.
 Reid-Walls-like chained inequalities are violated for values of gain $\Gamma$ decreasing with growing $L$. Surprisingly the behavior  of the ratios-based inequalities is opposite!. Table \ref{tab1} lists the critical values of  $\Gamma_{\text{R-W}}$ and $\Gamma_{\text{ratios}}$, below which the chained inequalities of ``length" $L$ are violated.
\begin{table}
	\centering
		\begin{tabular}{|c|c|c|c|c|c|}
		\hline
		$L$&$\Gamma_{\text{R-W}}^{crit}$&$\Gamma_{\text{ratios}}^{crit}$&$L$&$\Gamma_{\text{R-W}}^{crit}$&$\Gamma_{\text{ratios}}^{crit}$\\
		2&0.491&1.229&8&0.251&1.425\\
		3&0.408&1.342&9&0.237&1.431\\
		4&0.355&1.372&10&0.224&1.437\\
		5&0.318&1.392&11&0.214&1.441\\
		6&0.290&1.406&12&0.205&1.445\\
		7&0.268&1.416&13&0.197&1.448\\
		\hline			
		\end{tabular}
	\caption{Critical values of $\Gamma_{\text{R-W}}$ and $\Gamma_{\text{ratios}}$ above which the Reid-Walls-like and ratios-based chained Bell inequalities cannot be violated.}
	\label{tab1}
\end{table}

The new method gives improvement also in the case of chained inequalities recently derived in  Ref. \cite{CHAINED}.
The inequalities, if one uses the wording of this paper,  are based on averages of  $|I_{A_+}(\theta)-I_{B_+}(\phi)|$, which can be shown to have properties of geometric distance.
Let us replace this by $\langle|R_{A_+}(\theta)-R_{B_+}(\phi)|\rangle $. The resulting inequalities read 
\begin{eqnarray}
&\sum_{i=1}^L\left\langle\left|R_{A_+}(\theta_i)-R_{B_+}(\phi_i)\right|\right\rangle&    \nonumber\\
&+\sum_{i=1}^{L-1}  \langle\left|R_{A_+}(\theta_{i+1})-R_{B_+}(\phi_i)\right|\rangle &    \nonumber\\
&\geq\langle\left|R_{A_+}(\theta_1)-R_{B_+}(\phi_L)\right|\rangle.&
\end{eqnarray}

We have numerically studied the violation of the chained inequalities with the cutoff of at most 25 pairs produced in each run. The values of the gain, below which we still observe the violation of the inequalities are presented in Table \ref{tab2}. As expected, chained inequalities for ratios are violated for higher gains $\Gamma$ than the original ones. A simple explanation is that in the latter, the the contributions from high numbers of produced  pairs can grow unboundedly, while in the ratios-based version, they cannot be higher than 2.
Notice, for example, that to violate the inequality for ratios at $\Gamma=1.4$ one needs only 4 local settings, whereas it takes 7 settings to violate the original inequality.
\begin{table}
	\centering
		\begin{tabular}{|c|c|c|}
		\hline
		$L$&$\Gamma_{\text{Chained}}^{crit}$&$\Gamma_{\text{Chained, ratios}}^{crit}$\\
		\hline
		2&0.915&1.123\\
		3&1.053&1.367\\
		4&1.165&1.482\\
		5&1.260&1.586\\
		6&1.345&1.687\\
		7&1.427&1.795\\
		\hline			
		\end{tabular}
  \caption{Critical values of gain $\Gamma_{\text{Chained}}^{crit}$ and $\Gamma_{\text{Chained, ratios}}^{crit}$ above which chained inequalities for moduli of intensity differences and and similar ones for ratios cannot violated in function of the number of local settings $L$}
	\label{tab2}
\end{table}

\section{Closing remarks}
One can generalize the approach to situations involving more observation stations, compare Ref \cite{LASK}, and Bell experiments involving generalizations of Bell inequalities for many parties, like \cite{ZB}.

The new approach in fact leads to a possibility of a new look at Stokes parameters for non-classical light. This will be presented in a forthcoming paper \cite{ZUK}, where it will be shown that with the new approach we can get better entanglement witnesses (conditions). This is done by replacing the Stokes parameters in old entanglement conditions of the type presented in Ref. \cite{SIMON, MASZA, STOB-MASZA} by 
\begin{eqnarray} 
 \hat{S}'_{A}(\theta)
= \hat{\Pi}_A^{\perp 0}\frac{a^\dagger_\theta a_\theta  -  a^\dagger_{\theta_\perp} a_{\theta_\perp}    }{a^\dagger_\theta a_\theta+a^\dagger_{\theta_\perp} a_{\theta_\perp}}  \hat{\Pi}_A^{\perp 0}.
\end{eqnarray}
As a matter of fact all entanglement witnesses for states of two-qubit systems can be directly mapped into entanglement indicators, for optical four mode fields (of the kind studied here),  involving polarization measurements.

\section{Acknowledgments} 
The work is a part of  EU grant BRISQ2.  
It is subsidized form funds for science for years 2012-2015 approved for
international co-financed project BRISQ2 by Polish Ministry of Science and Higher Education (MNiSW).
MZ and WL are supported by TEAM project of FNP. 
MW acknowledges the financial support of the Polish National
Science Centre (NCN), grant DEC-2013/11/D/ST2/02638.
We thank prof. M. Chekhova for discussions.

%CZY UMIECIE LICZYC SLOWA W LATEX'U?

\end{document}